\documentclass[aps,preprint,nofootinbib,showpacs]{revtex4}
\usepackage{graphicx,color,amsmath, epsfig, rotating}
\newcommand{\be}{\begin{equation}}
\newcommand{\ee}{\end{equation}}
\newcommand{\ba}{\begin{eqnarray}}
\newcommand{\beq}{\begin{equation}}
\newcommand{\eeq}{\end{equation}}
\newcommand{\ea}{\end{eqnarray}}

\newcommand{\tr}{{\rm Tr}}

\def\beqa{\begin{eqnarray}}
\def\eeqa{\end{eqnarray}}

\def\bea{\begin{eqnarray}}
\def\eea{\end{eqnarray}}

\def\err#1#2{\lower2pt\hbox{ $\stackrel{\scriptstyle +#1}{\scriptstyle -#2}$}}
\def\ga{\mathrel{\raise.3ex\hbox{$>$\kern-.75em\lower1ex\hbox{$\sim$}}}}
\def\la{\mathrel{\raise.3ex\hbox{$<$\kern-.75em\lower1ex\hbox{$\sim$}}}}
\def\bmaT{\left(\begin{array}{ccc}}
\def\emaT{\end{array}\right)}
\def\bma{\left( \begin{array} }
\def\ema{\end{array} \right)}
\def\gsim{~{\rlap{\lower 3.5pt\hbox{$\mathchar\sim$}}\raise 1pt\hbox{$>$}}\,}
\def\lsim{~{\rlap{\lower 3.5pt\hbox{$\mathchar\sim$}}\raise 1pt\hbox{$<$}}\,}

\begin{document}

\title{\boldmath Phenomenology of Large Mixing for the CP-even Neutral \\Scalars
of the Higgs Triplet Model\unboldmath} 

\author{A.G. Akeroyd\footnote{akeroyd@ncu.edu.tw}}
\affiliation{Department of Physics and Center for Mathematics and Theoretical
  Physics, National Central University, Chungli, Taiwan 320}
\author{Cheng-Wei Chiang\footnote{chengwei@ncu.edu.tw}}
\affiliation{Department of Physics and Center for Mathematics and Theoretical
  Physics, National Central University, Chungli, Taiwan 320}
\affiliation{Institute of Physics, Academia Sinica, Taipei, Taiwan 115}

\date{\today}

\begin{abstract}
The Higgs Triplet Model contains two CP-even neutral scalar eigenstates, each having components from an isospin doublet and an isospin triplet scalar field. The mixing angle can be maximal if the masses of the scalar eigenstates are close to degeneracy. We quantify the dependence of the mixing angle on the mass splitting and on the vacuum expectation value of the neutral triplet scalar. We 
determine the parameter space for maximal mixing, and study the observability 
of both CP-even Higgs bosons at the CERN LHC.
\end{abstract}

\maketitle


\section{Introduction}

The firm evidence that neutrinos oscillate and possess small masses below the eV scale \cite{Fukuda:1998mi} necessitates physics beyond the Standard Model (SM), which could manifest itself at the CERN Large Hadron Collider (LHC) and/or in low energy experiments which search for lepton flavour violation (LFV) \cite{Kuno:1999jp}.  Consequently, models of neutrino mass generation which can be probed at present and forthcoming experiments are of great phenomenological interest.

Neutrinos may obtain masses via the vacuum expectation value (VEV) of a neutral Higgs boson in an isospin triplet representation \cite{Konetschny:1977bn, Mohapatra:1979ia, Magg:1980ut, Schechter:1980gr,Cheng:1980qt}.  A particularly simple implementation of this mechanism of neutrino mass generation is the ``Higgs Triplet Model'' (HTM) in which the SM Lagrangian is augmented solely by an $SU(2)$ triplet of scalar particles with hypercharge $Y=2$~\cite{Konetschny:1977bn, Schechter:1980gr,Cheng:1980qt}.  In the HTM, neutrinos acquire Majorana masses given by the product of a triplet Yukawa coupling ($h_{ij}$) and a triplet VEV ($v_\Delta$).  Consequently, there is a direct connection between $h_{ij}$ and the neutrino mass matrix, which gives rise to phenomenological predictions for processes which depend on $h_{ij}$ \cite{Ma:1998dx,Chun:2003ej,Kakizaki:2003jk,Garayoa:2007fw,Akeroyd:2007zv,%
  Kadastik:2007yd,Perez:2008ha,delAguila:2008cj,Akeroyd:2009nu,Fukuyama:2009xk,%
  Akeroyd:2009hb}.  A distinctive signal of the HTM would be the observation of a doubly charged Higgs boson ($H^{\pm\pm}$), whose mass ($M_{H^{\pm\pm}}$) may be of the order of the electroweak scale. Such particles can be produced with sizeable rates at hadron colliders in the processes $q\overline q\to H^{++}H^{--}$ \cite{Barger:1982cy,Gunion:1989in,Han:2007bk,Huitu:1996su} and $q\overline {q'}\to H^{\pm\pm}H^{\mp}$~\cite{Barger:1982cy,Dion:1998pw, Akeroyd:2005gt}, where $H^\pm$ is a singly charged Higgs boson in the same triplet representation.  Direct searches for $H^{\pm\pm}$ have been carried out at the Fermilab Tevatron, assuming the production channel $q\overline q\to H^{++}H^{--}$ and decays $H^{\pm\pm}\to \ell^\pm_i\ell^\pm_j$, and mass limits in the range $M_{H^{\pm\pm}}> 110\to 150$~GeV have been obtained \cite{Acosta:2004uj,Abazov:2004au,:2008iy,Aaltonen:2008ip}.  The CERN Large Hadron Collider (LHC), using the above production mechanisms, will offer improved sensitivity to $M_{H^{\pm\pm}}$ \cite{Perez:2008ha,delAguila:2008cj,Han:2007bk,Hektor:2007uu}.  The phenomenology of the singly charged Higgs boson is also attractive at hadron colliders, with production via $q\overline {q'}\to H^{\pm\pm}H^{\mp}$ followed by the decay $H^\pm\to \ell^\pm\nu$ \cite{Perez:2008ha,delAguila:2008cj,Akeroyd:2009hb}.

The phenomenology of the neutral Higgs bosons in the HTM has received much less attention than that of the charged Higgs bosons. There are two CP-even scalars ($H_1,H_2$, where $M_{H_2} > M_{H_1}$) and one CP-odd scalar ($A^0$), which are composed of both isospin doublet and isospin triplet fields. In phenomenological studies of the HTM, it is common to assume that the mass term for the scalar triplet ($\sim M_\Delta$) is considerably larger than the mass of the isospin doublet scalar.  This assumption guarantees that the mixing angle for the two CP-even scalars is small, being of the order $v_{\Delta}/v_0$ ($v_0$ is the VEV of the isospin doublet), where $v_{\Delta} \ll v_0$ is required to maintain $\rho\equiv M^2_W/(M^2_Z\cos^2\theta_W) \sim 1$ within experimental error. Therefore, $H_1$ is essentially composed of the isospin doublet field and plays the role of the SM Higgs boson, while $H_2$ is essentially composed of the isospin triplet field, and is difficult to detect at hadron colliders.

However, as pointed out explicitly in Ref.~\cite{Dey:2008jm}, the mixing angle for the CP-even scalars can be maximal in the region of parameter space around degenerate masses for the CP-even scalars. We quantify in detail this parameter space of large mixing in the CP-even sector, and study its phenomenology. When the mixing angle is large, $H_2$ can be produced with observable rates in the standard search channels for the SM Higgs boson. Importantly, $H_2$ in the HTM can be considerably lighter than $H^{\pm\pm}$ and $H^{\pm}$.  Therefore, $H_2$ might be detected earlier than $H^{\pm\pm}$ or $H^{\pm}$, especially if the decay modes $H^{\pm\pm}\to W^\pm W^\pm$ and $H^{\pm}\to W^\pm Z$ dominate (corresponding to $v_\Delta > 10^{-3}$ GeV), for which the LHC has sensitivity inferior to that for the leptonic channels $H^{\pm\pm}\to \ell^\pm_i\ell^\pm_j$ and $H^\pm\to \ell^\pm\nu$ described above.
 
Our work is organized as follows.  The HTM is briefly reviewed in Section II.  In Section III, the scalar mass matrices, the mixing angle, and the Higgs potential minimization and stability conditions are presented.  The numerical analysis and phenomenology are discussed in Section IV.  Our conclusions are given in Section V.

\section{The Higgs Triplet Model}

In the HTM, an $I=1,Y=2$ complex $SU(2)_L$ triplet of scalar fields is added to the SM Lagrangian.  Such a model can provide a Majorana mass for the observed neutrinos (without the introduction of additional neutrinos) via the $SU(2)\otimes U(1)_Y$
 gauge-invariant Yukawa interactions:
\begin{equation}
{\cal L}=h_{ij}\psi_{iL}^TCi\sigma_2\Delta\psi_{jL} + {\rm h.c.}
\label{trip_yuk}
\end{equation}
Here $h_{ij} (i,j=e,\mu,\tau)$ is a complex and symmetric coupling matrix, $C$ is the Dirac charge conjugation operator, $\sigma_2$ is a Pauli matrix, $\psi_{iL}=(\nu_i, \ell_i)_L^T$ is a left-handed lepton doublet, and $\Delta$ is a $2\times 2$ representation of the $Y=2$ complex triplet fields:
\begin{equation}
\Delta
=\bma{cc}
\delta^+/\sqrt{2}  & \delta^{++} \\
\delta^0       & -\delta^+/\sqrt{2}
\ema ~.
\end{equation}
A non-zero triplet VEV, $\langle\delta^0\rangle=v_\Delta/\sqrt 2$, gives rise to the following mass matrix for neutrinos:
\begin{equation}
m_{ij}=2h_{ij}\langle\delta^0\rangle = \sqrt{2}h_{ij}v_{\Delta} ~.
\label{nu_mass}
\end{equation}
This simple expression for tree-level Majorana masses of the observed neutrinos is essentially the main motivation for studying the HTM.  Realistic neutrino masses can be obtained with a perturbative $h_{ij}$ provided that $v_{\Delta}\gsim 1$ eV.  The presence of a non-zero $v_{\Delta}$ gives rise to $\rho\ne 1$ at tree level.  Therefore $v_{\Delta} \lsim 1$ GeV is necessary in order to comply with the measurement of $\rho\sim 1$.  We will discuss this bound on $v_{\Delta}$ in more detail later.

Neutrino oscillation experiments have provided much information on $m_{ij}$ (see, for example, Ref.~\cite{Schwetz:2008er}), and so the couplings $h_{ij}$ are already constrained up to an arbitrary scalar factor (the triplet VEV, $v_\Delta$). The necessary non-zero $v_{\Delta}$ arises from the minimization of the most general $SU(2)\otimes U(1)_Y$ invariant Higgs potential, which is written as follows \cite{Joshipira:1991yy, Ma:1998dx, Perez:2008ha}:
\begin{eqnarray}
V(H,\Delta) & = & 
- m_H^2 \ H^\dagger H \ + \ \frac{\lambda}{4} (H^\dagger H)^2 \ 
+ \ M_{\Delta}^2 \ \tr \Delta^\dagger \Delta\ 
+ \ \left( \mu \ H^T \ i \sigma_2 \ \Delta^\dagger H \ + \ {\rm h.c.}\right) \ 
\nonumber \\
&& + \ \lambda_1 \ (H^\dagger H) \tr \Delta^\dagger \Delta \ 
+ \ \lambda_2 \ \left( \tr \Delta^\dagger \Delta \right)^2 \ 
+ \ \lambda_3 \ \tr \left( \Delta^\dagger \Delta \right)^2 \ 
+ \ \lambda_4 \ H^\dagger \Delta \Delta^\dagger H ~.
\label{Potential}
\end{eqnarray}
Here $H=(\phi^+,\phi^0)^T$ is the SM Higgs doublet.  Variants of the above form for $V(H,\Delta)$ are given in Refs.~\cite{Chun:2003ej, Dey:2008jm, Abada:2007ux,Gogoladze:2008gf}, which are equivalent to a reparametrization of some $\lambda_i$.

As in the SM, the term $-m^2_H$ (where $m^2_H>0$) ensures $\langle\phi^0\rangle=v_0/\sqrt 2$, which spontaneously breaks
 $SU(2)\otimes U(1)_Y$ to $U(1)_Q$. The mass term for the triplet
 scalars is given by $M^2_{\Delta}$ and usually $M^2_{\Delta}>0$ is taken. 
One can take values of $M^2_{\Delta}$ which are arbitrarily large,
 but recently much attention has been given to the case of $M_{\Delta} < 1$ TeV,
 since this would allow the triplet scalars (especially the distinctive doubly charged scalar, $H^{\pm\pm}$) to be within the discovery range of the LHC. The main production mechanisms for $H^{\pm\pm}$ at hadron colliders are (i) $q\overline q\to H^{++}H^{--}$ \cite{Barger:1982cy,Gunion:1989in,Han:2007bk,Huitu:1996su}, which depends on one unknown parameter, $M_{H^{\pm\pm}}$; and (ii) $qq'\to H^{\pm\pm}H^{\mp}$~\cite{Barger:1982cy, Dion:1998pw, Akeroyd:2005gt}, which depends on two unknown parameters, $M_{H^{\pm\pm}}$ and $M_{H^{\pm}}$. In the HTM one has $M_{H^{\pm\pm}}\sim M_{H^{\pm}}$ if $\lambda_4$ is small (see later). Production mechanisms which depend on the triplet VEV ($pp\to W^{\pm *}\to W^\mp H^{\pm\pm}$ and fusion via $W^{\pm *} W^{\pm *} \to H^{\pm\pm}$ \cite{Huitu:1996su, Vega:1989tt}) are not competitive with the above processes at the energies of the Fermilab Tevatron, but can be the dominant source of $H^{\pm\pm}$ at the LHC if $v_{\Delta}={\cal O}$ (1 GeV) and 
$M_{H^{\pm\pm}}> 500$ GeV.

The term $\mu(\Phi^Ti\sigma_2\Delta^\dagger\Phi)$ leads to the triplet VEV, as will be shown explicitly in the next section.  In an early version of the HTM \cite{Gelmini:1980re}, the term $\mu(\Phi^Ti\sigma_2\Delta^\dagger\Phi)$ is absent, but $v_\Delta$ can still arise by taking the ``wrong sign'' choice for $M^2_{\Delta}$ ($<0$). This leads to spontaneous violation of the lepton number since Majorana mass has come from a Higgs potential which originally conserves the lepton number. The resulting Higgs spectrum then contains a massless triplet scalar (called Majoron, $J$, a Goldstone boson) and another light scalar ($H^0$). This is a dramatic prediction, and pair production via $e^+e^-\to Z\to H^0J$ would give a large contribution to the invisible width of $Z$. This model is therefore testable, and it is now excluded because the invisible width for $Z$ has been measured at the CERN Large Electron Positron Collider (LEP), and its value is in good agreement with the SM prediction (in which the invisible width comes from $Z\to \nu_i\overline \nu_i$ only).
 
The inclusion of the term $\mu(\Phi^Ti\sigma_2\Delta^\dagger\Phi$) explicitly breaks lepton number when $\Delta$ is assigned $L=2$, and eliminates the Majoron. Alternatively, assigning $L=0$ would conserve lepton number in the Higgs potential but break it in the Yukawa interaction of Eq.~(\ref{trip_yuk}).  Therefore, the lepton number is always broken irrespective of the assignment of $L=0$ or $L=2$ because of the presence of both Eq.~(\ref{trip_yuk}) and $\mu(\Phi^Ti\sigma_2\Delta^\dagger\Phi$).  Thus the above scalar potential together with the triplet Yukawa interactions of Eq.~(\ref{trip_yuk}) lead to a model of neutrino mass generation which is viable phenomenologically.

One can work in a simplified scalar potential (e.g., Ref.~\cite{Perez:2008ha}) by neglecting the quartic couplings $\lambda_i$ (where $i=1,2,3,4$) involving the triplet field $\Delta$.  The resulting scalar potential then depends on four parameters ($-m^2_H$, $\lambda$, $\mu$, $M_{\Delta}$), but only three parameters are independent because the VEV for the doublet field ($v_0=246$ GeV) is fixed by the mass of $W^\pm$. The three independent parameters are usually chosen as $\lambda,\mu,M_{\Delta}$ or $\lambda,\mu,v_\Delta$.  The inclusion of $\lambda_i$ generates additional trilinear and quartic couplings among the scalar mass eigenstates, which contribute to the term which mixes the CP-even scalars.  The terms with $\lambda_1$ and $\lambda_4$, which involve both triplet and doublet fields, are of particular interest because they can give sizeable contributions to the masses of the triplet fields (when replacing the fields $H^\dagger H$ by $v_0^2$).  In this work we will study the HTM with a scalar potential given by Eq.~(\ref{Potential}).

\section{Minimization equations and mass matrices for the scalar fields}

Following the notation of Ref.~\cite{Perez:2008ha}, the neutral complex scalar fields are expressed as follows:
\begin{equation} 
\phi^0 = ( v_0 \ + \ h^0 \ + i \xi^0 )/ \sqrt{2} ~, ~~ \text{and}~~
\delta^0 = ( v_{\Delta} \ + \ \Delta^0 \ + i \eta^0)/\sqrt{2} ~.
\label{SMhiggs}
\end{equation}
For non-zero $v_0$ and $v_\Delta$, the minimization conditions for the global minimum of the potential are:
\begin{eqnarray}
&&
-m_H^2 + \frac{\lambda}{4}v_0^2 + \frac12 (\lambda_1 + \lambda_4) v_\Delta^2 -
\sqrt{2} \mu v_\Delta = 0 ~, \mbox{ and} \\
&&
M_\Delta^2 v_\Delta + \frac12 (\lambda_1 + \lambda_4) v_0^2 v_\Delta
- \frac{1}{\sqrt{2}} \mu v_0^2 + (\lambda_2 + \lambda_3) v_\Delta^3 = 0 ~.
\end{eqnarray}
In the simplified potential which sets $\lambda_i=0$, the expression for 
$v_\Delta$ resulting from the minimization of $V$ is:
\begin{equation}
v_\Delta = \frac{\mu v_0^2}{\sqrt 2 M^2_{\Delta}} \ .
\label{tripletvev}
\end{equation}
For $M_{\Delta} \gg v_0$, this expression is sometimes referred to as the ``Type II seesaw mechanism,'' since a small value for $v_\Delta$ arises without requiring a small value of $\mu$.  However, the case of $M_{\Delta} < 1$ TeV is of phenomenological interest because the triplet scalars would be produced at the LHC, and such a scenario requires $\mu \sim v_\Delta <$ a few GeV.

After imposing the above tadpole conditions to eliminate $m_H$ and $M_\Delta$, one finds that the mass-squared matrix ($\frac{1}{2}[h^0,\Delta^0]{\cal M}_{\rm even}^2[h^0,\Delta^0]^T$) for the CP-even states is:
\begin{eqnarray}
{\cal M}^2_{\rm even} = 
\left(\begin{array}{cc}
\lambda v_0^2 / 2 &
\left[ (\lambda_1 + \lambda_4) v_\Delta - \sqrt{2} \mu \right] v_0 \\
\left[ (\lambda_1 + \lambda_4) v_\Delta - \sqrt{2} \mu \right] v_0 &
(\sqrt{2}\mu v_0^2 + 4(\lambda_2 + \lambda_3)v_\Delta^3) / 2v_\Delta
\end{array}\right) ~.
\label{CP-even}
\end{eqnarray}
Note that
${\cal M}^2_{\rm even}$ depends on all seven parameters of the scalar potential. 
The mass eigenstates are denoted by $H_1$ and $H_2$, where $M_{H_2}>M_{H_1}$:
\begin{eqnarray}
H_1 &=& \cos \theta_0 \ h^0 \ + \ \sin \theta_0 \ \Delta^0,
\quad
H_2 = - \sin \theta_0 \ h^0 \ + \ \cos \theta_0 \Delta^0.
\qquad
\label{CP-even-mix}
\end{eqnarray}

The square of the mass eigenvalues $M_{H_1},M_{H_2}$ are given by:
\begin{equation}
M^2_{H_1},M^2_{H_2} = \frac{1}{2}
\left[ {\cal M}^2_{11}+{\cal M}^2_{22}\pm \sqrt{({\cal M}^2_{11}-{\cal M}^2_{22})^2+
(4{\cal M}^2_{12})^2} \right]
\label{eigenvalues}
\end{equation}
For the case of ${\cal M}^2_{22}>{\cal M}^2_{11}$, the explicit expressions for the squared masses of $H_1$ and $H_2$ expanded to terms linear in $\epsilon=v_\Delta/v_0$ are:
\begin{eqnarray}
M^2_{H_1}= 
\frac{1}{2}\lambda v^2_0-2\sqrt 2\mu v_0\epsilon + {\cal O}(\epsilon^2) ~, \\
M^2_{H_2}= 
\frac{\mu v_0}{\sqrt 2\epsilon}+2\sqrt 2\mu v_0\epsilon + {\cal O}(\epsilon^2) ~.
\label{CPeven-masses}
\end{eqnarray}

The mass-squared matrix ($\frac{1}{2}[\xi^0,\eta^0]{\cal M}_{\rm odd}^2[\xi^0,\eta^0]^T$) for the CP-odd states is:
\begin{eqnarray}
{\cal M}^2_{\rm odd} = \mu
\left(\begin{array}{cc}
2\sqrt{2} v_\Delta & -\sqrt{2} v_0 \\
-\sqrt{2}  v_0 &  v_0^2 / (\sqrt{2} v_\Delta)
\end{array}\right) ~,
\end{eqnarray}
which is completely independent of the scalar quartic couplings ($\lambda_i$ and $\lambda$) and only depends on two parameters ($\mu$ and $v_\Delta$) of the scalar potential. One of the eigenstates is the neutral Goldstone boson which becomes the longitudinal polarization mode of the $Z$ boson after the breaking of the $SU(2)\otimes U(1)_Y$ symmetry.  The massive eigenstate is denoted by $A^0$:
\begin{equation}
A^0 = - \sin \alpha \ \xi^0 \ + \ \cos \alpha \ \eta^0.\qquad
\label{CP-odd}
\end{equation}
The squared mass of $A^0$ is given by:
\begin{equation}
M^2_{A^0} = \frac{\mu v_0}{\sqrt 2 \epsilon} + 2\sqrt2 \mu v_0 \epsilon \qquad
\label{CP-odd-mass}
\end{equation}
Note that $A^0$ also becomes massless (``a triplet Majoron'') in the limit of $\mu\to 0$ \cite{Gelmini:1980re} and $v_\Delta\ne 0$, i.e., the scenario of spontaneous breaking (not explicit breaking) of lepton number caused by $M^2_\Delta<0$. For our choice of $M^2_\Delta>0$ and $\mu\ne 0$, the sign of $\mu$ and $v_\Delta$ must be the same in order to ensure a positive mass for $A^0$. We choose $\mu$ and $v_\Delta$ to be positive.

The mass-squared matrix for the singly charged states ($[\phi^+,\delta^+]{\cal M}_{\pm}^2[\phi^-,\delta^-]^T$)
is:
\begin{eqnarray}
{\cal M}^2_{\pm} = \left( \mu-\frac{\lambda_4 v_\Delta}{2\sqrt 2} \right)
\left(\begin{array}{cc}
\sqrt{2} v_\Delta &
-v_0  \\
-v_0  &
v_0^2/(\sqrt{2} v_\Delta)
\end{array}\right) ~,
\end{eqnarray}
where one of the eigenstates has a vanishing eigenvalue and serves as the charged Goldstone boson that later becomes the longitudinal polarization mode of the $W$ boson. The massive eigenstate is denoted by $H^\pm$:
\begin{equation}
H^{\pm} = - \sin \theta_{\pm} \ \phi^{\pm} \ + \ \cos \theta_{\pm} \ \delta^{\pm}
~.
\label{charged}
\end{equation}
Note that only one scalar quartic coupling ($\lambda_4$) appears in ${\cal M}^2_{\pm}$, and the mass matrix depends on three parameters.  The squared mass of $H^\pm$ is given by:
\begin{equation}
M^2_{H^\pm} = 
\frac{\mu v_0}{\sqrt2 \epsilon} - \frac{\lambda_4}{4} v_0^2 + \sqrt 2 \mu v_\Delta
-\frac{\lambda_4}{2} v_\Delta^2 ~.
\label{charged-mass}
\end{equation}

Finally, the squared mass of the doubly-charged state ($H^{\pm\pm} = \delta^{\pm\pm}$) is given by:
\begin{equation}
M^2_{H^{\pm\pm}} =
\frac{\mu v_0^2}{\sqrt{2} v_\Delta} - \frac{\lambda_4}{2} v_0^2
-\lambda_3 v_\Delta^2 ~,
\label{doub-charged-mass}
\end{equation}
which depends on four parameters of the model.

The above mass matrices for the neutral scalar fields are presented in Ref.~\cite{Joshipira:1991yy} in the context of an extension of the HTM which includes a singlet scalar field.  The mass matrices for the neutral and charged scalars are given in Ref.~\cite{Ma:1998dx} in the approximation of 
neglecting one or two of $\lambda_i$ (see also Ref.~\cite{Frampton:2002rn}).
The scalar mass matrices for the Majoron model with $M^2_\Delta<0$ 
and $\mu=0$ are given in Ref.~\cite{Gelmini:1980re, Georgi:1981pg}.

\begin{figure}[h]
\begin{center}
\includegraphics[origin=c, angle=0, scale=0.50]{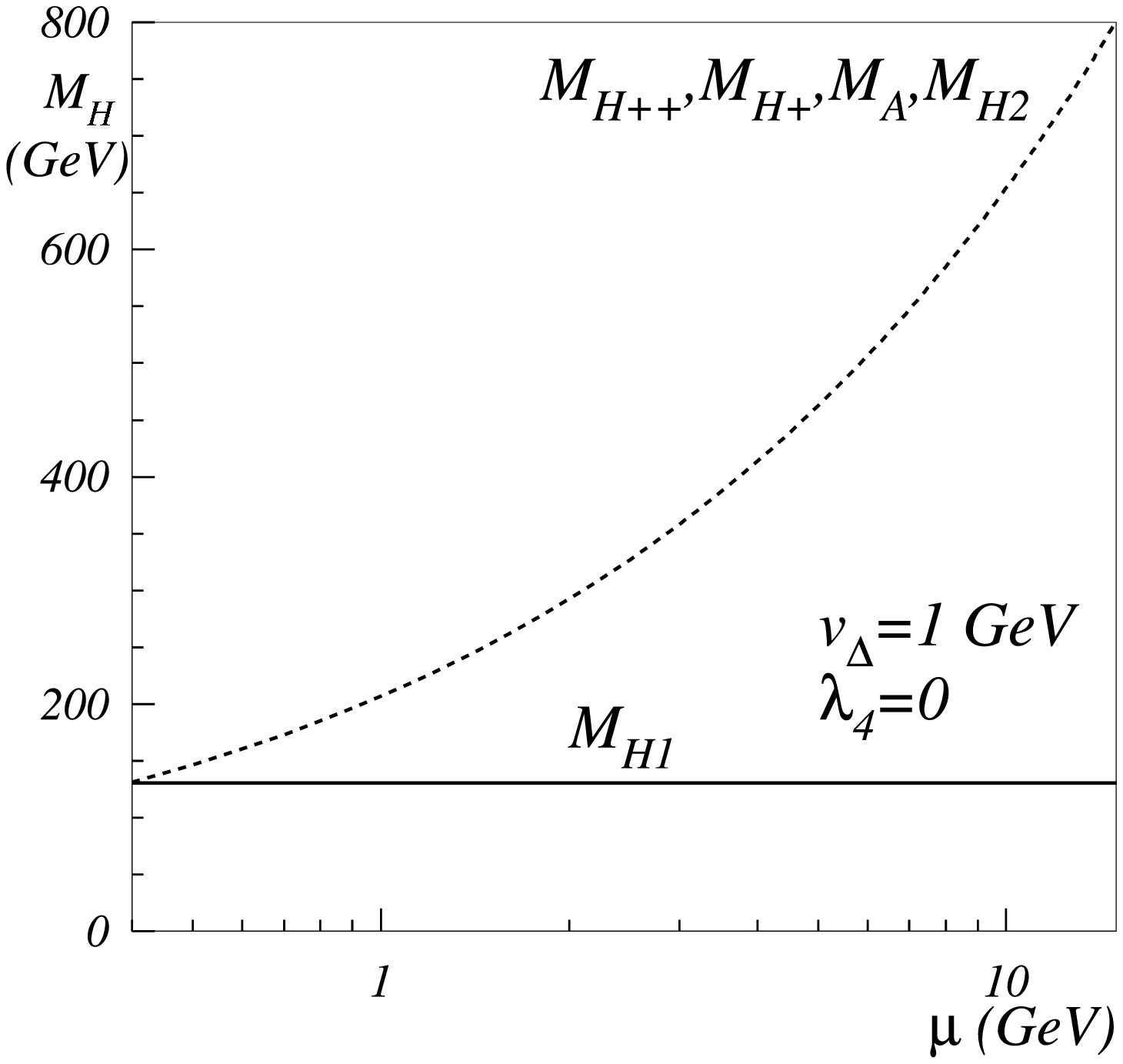}
\includegraphics[origin=c, angle=0, scale=0.50]{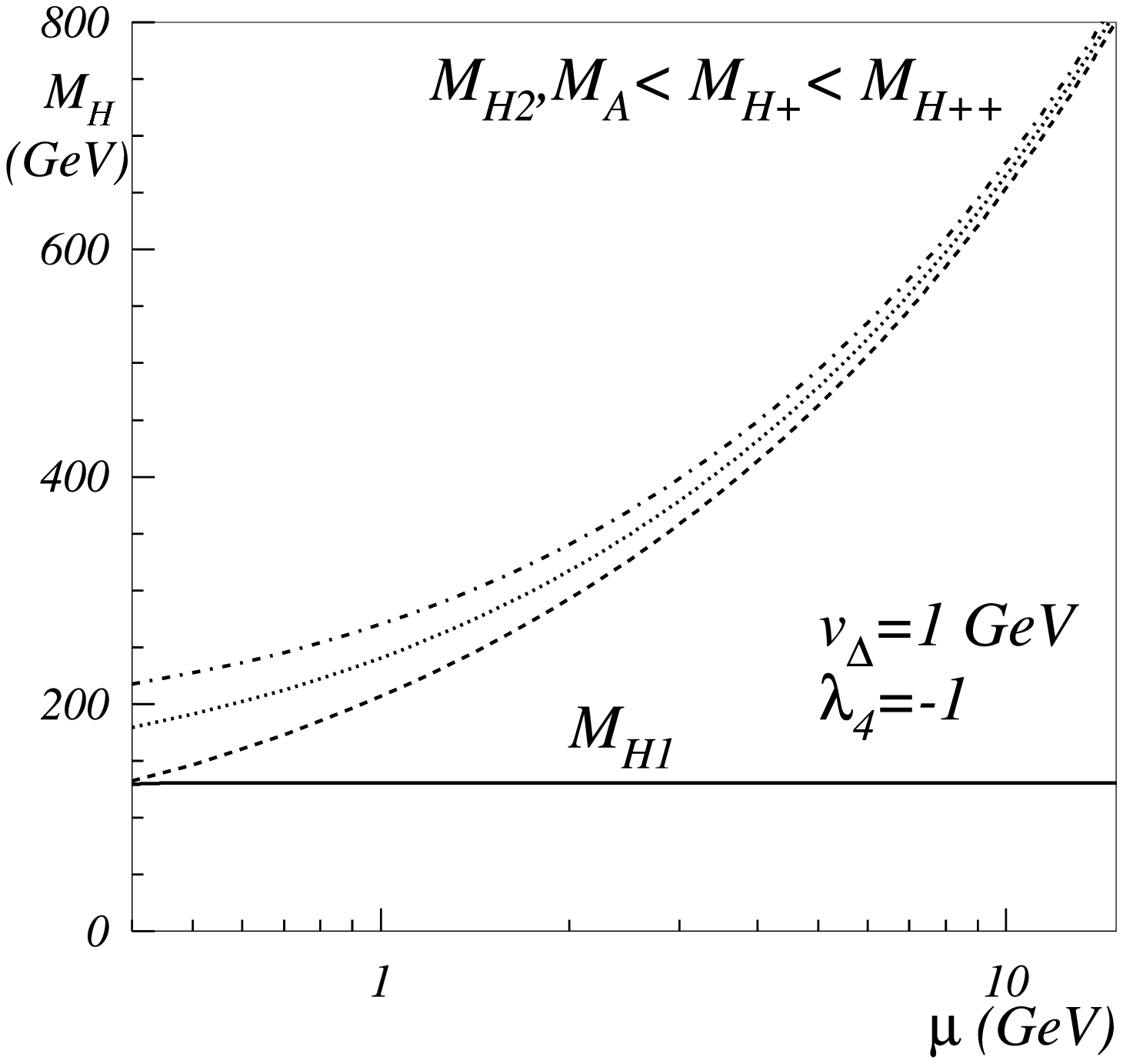}
\includegraphics[origin=c, angle=0, scale=0.50]{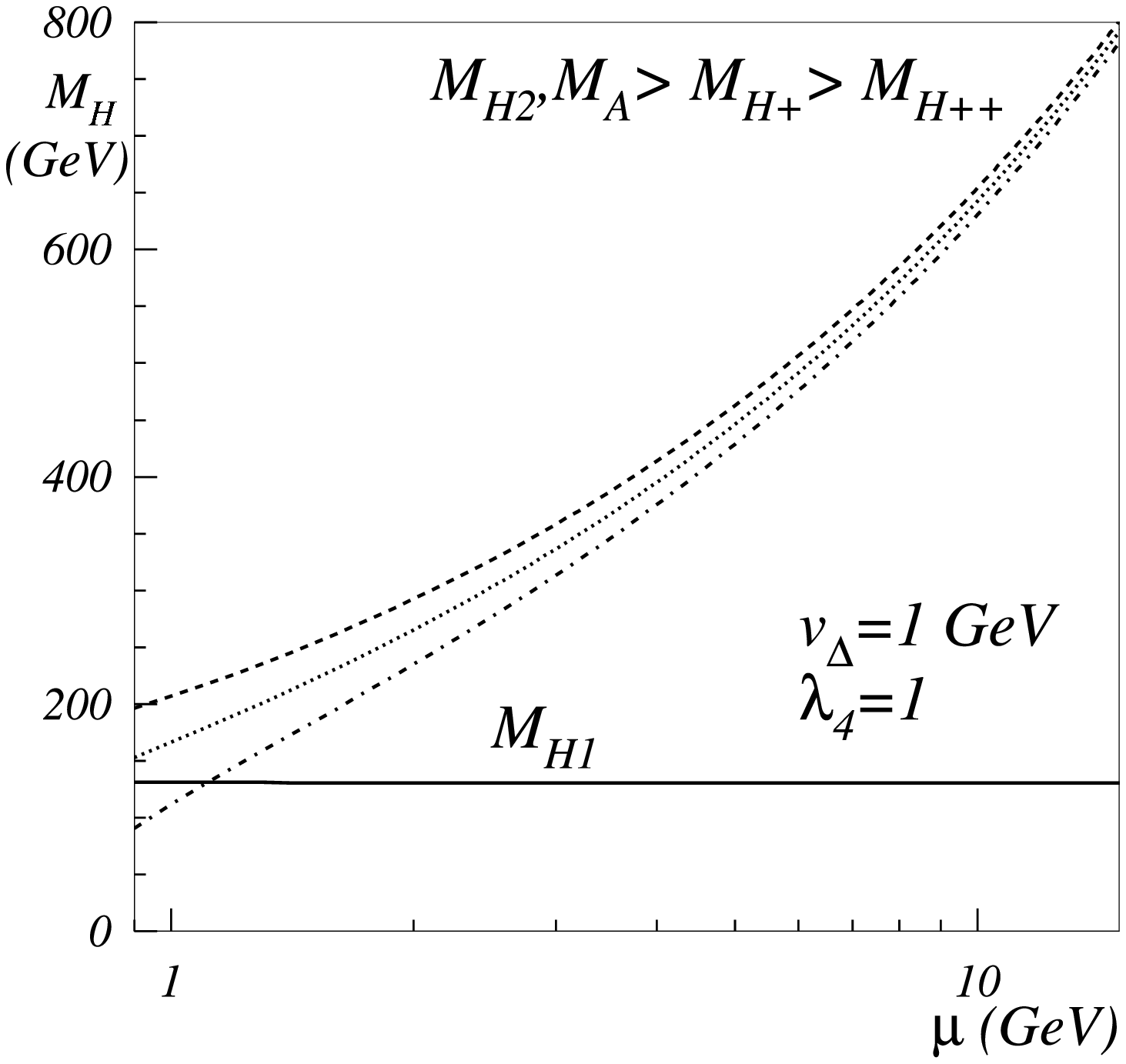}
\vspace*{10mm}
\caption{The masses of the Higgs bosons ($M_{H^{\pm\pm}}$, $M_{H^{\pm}}$, $M_{A^0}$, $M_{H_1}$ and $M_{H_2}$) as a function of $\mu$. The upper panels correspond to 
$\lambda_4=0$ (left) and $\lambda_4=-1$ (right), and the lower panel corresponds to $\lambda=1$.  Other parameters are fixed as follows: triplet VEV $v_\Delta=1$ GeV, $\lambda \simeq 0.566$, $\lambda_1=0$, and $\lambda_{2,3}=1$.}
\label{fig:masses}
\end{center}
\end{figure}

In Fig.~\ref{fig:masses}, the masses of the scalars of the HTM are presented as a function of $\mu$ for three values of $\lambda_4$ ($=0,1$ and $-1$).  Other parameters are fixed as follows: triplet VEV $v_\Delta=1$ GeV, $\lambda\simeq0.566$, $\lambda_1=0$, and $\lambda_{2,3}=1$. The current experimental bounds on the masses of the scalars are respected by choosing $\mu$ above a threshold value. In the plots for $\lambda_4=0$ and $\lambda_4=-1$ we take $\mu>0.4$, while in the plot for $\lambda_4=1$ we take $\mu>0.9$.  We note that the mass of the doubly charged scalar starts at $M_{H^{\pm\pm}}\simeq 90$ GeV in the plot with $\lambda_4=1$. This is not in conflict with the experimental lower bound because the decay mode $H^{\pm\pm}\to W^\pm W^\pm$ has a branching ratio $\simeq 100\%$ for our choice of $v_\Delta=1$ GeV, and there has been no explicit search for $H^{\pm\pm}$ in this decay channel.  As $\mu$ increases (or equivalently, as $M_\Delta$ increases) one can see that $M_{H_1}$ remains constant, its magnitude being determined by the entry ${\cal M}^2_{11}$ in the CP-even mass matrix ${\cal M}^2_{\rm even}$, which is independent of $\mu$.  In contrast, $M_{H^{\pm\pm}},M_{H^{\pm\pm}},M_{A^0}$ and $M_{H_2}$ all increase with $\mu$. The dominant part of the mass splitting among them is proportional to $\lambda_4 v_0^2$.  For the case of $\lambda_4=0$, the latter scalars are approximately degenerate, with very small splittings caused by electromagnetic corrections (see, for example, Ref.~\cite{Perez:2008ha}), and from other $\lambda_i$ which have a dependence on the small parameter $v_\Delta$ (see the explicit expressions for the masses of the scalars given in Eqs.~(\ref{CPeven-masses}), (\ref{CP-odd-mass}), (\ref{charged-mass}) and (\ref{doub-charged-mass})).  For $\lambda_4<0 \;(> 0)$ one has the mass hierarchy $M_{H^{\pm\pm}}> M_{{H^\pm}} > M_{{H_2,A^0}}$ ($M_{H^{\pm\pm}}< M_{{H^\pm}} < M_{{H_2,A^0}}$).  In Ref.~\cite{Dey:2008jm}, the analogous versions of Fig.~\ref{fig:masses} show a sizeable mass splitting between $M_{H_2}$ and the degenerate scalars $M_{H^{\pm\pm}}$, $M_{H^{\pm\pm}}$, $M_{A^0}$, even when $M_\Delta>> v_0$. We cannot reproduce this result.

In Fig.~\ref{fig:masses}, there is a region where $H_1$ and $H_2$ are approximately degenerate in mass.  This corresponds to the case of ${\cal M}^2_{11}\sim {\cal M}^2_{22}$ for the CP-even mass matrix, for which the mixing angle $\theta_0$ in Eq.~(\ref{CP-even-mix}) becomes maximal.  The mixing angles ($\theta_0,\alpha \,\,{\rm or}\,\, \theta_{\pm}$) for the mass matrices (${\cal M}^2_{\rm even}, {\cal M}^2_{\rm odd}, \,{\rm or}\, {\cal M}^2_{\pm}$) are given by the general formula:
\begin{equation}
\tan 2\theta=\frac{2{\cal M}^2_{12}}{{\cal M}^2_{11}-{\cal M}^2_{22}} ~.
\label{mixing_angle}
\end{equation}
Maximal mixing ($\theta=\pm 45^\circ$) is achieved at ${\cal M}^2_{11}={\cal M}^2_{22}$, irrespective of the value of ${\cal M}^2_{12}$ (provided that ${\cal M}^2_{12}\ne 0$).  The condition ${\cal M}^2_{11}={\cal M}^2_{22}$ is realized in each of the above mass matrices for the following choice of parameters, respectively:
\begin{eqnarray}
{\cal M}^2_{\rm even}: && \lambda=\sqrt 2\mu/v_\Delta + 4(\lambda_2+\lambda_3)\frac{v^2_\Delta}{v^2_0}
\label{maxmix}
\\
{\cal M}^2_{\rm odd}: && 4v^2_{\Delta}=v_0^2\\
{\cal M}^2_{\pm}: && 2v^2_{\Delta}=v_0^2
\end{eqnarray}
Maximal mixing can never be achieved for ${\cal M}^2_{\rm odd}$ and ${\cal M}^2_{\pm}$ because of the constraint $v_{\Delta}\lsim 1 \,{\rm GeV} \ll v_0$. Hence the mixing angles for the CP-odd and singly charged scalars are always small in the HTM, with $\tan2\alpha\sim \tan2\theta_{\pm} \sim 4v_{\Delta}/v_0$. However, maximal mixing is possible for the CP-even scalars and has been discussed first in Ref.~\cite{Dey:2008jm}. In Eq.~(\ref{maxmix}), if one neglects the small term proportional to $\lambda_2+\lambda_3$ (which is suppressed by $v^2_\Delta/v^2_0$) one has a simple condition for maximal mixing for ${\cal M}^2_{\rm even}$, given by $\lambda=\sqrt 2\mu/v_\Delta$.  This condition for $\lambda$ can be satisfied in the HTM, provided that the masses of the scalars respect their current lower bounds. We will study in detail the phenomenology of the scenario of a large mixing angle $\theta_0$ for the CP-even scalars, which is possible when $M_{H_1}$ and $M_{H_2}$ are approximately degenerate.

In general, the mixing angle for the CP-even scalars will be non-zero, apart from fine-tuned choices of parameters for which ${\cal M}^2_{12}=0$. In the limit of $M_{H_2} \gg M_{H_1}$ one has $\tan2\theta_0\sim 4v_\Delta/v_0$.  Notably, even a very small isospin doublet component (e.g., corresponding to values of $v_\Delta\sim {\rm 1\,MeV}$) can dominate the branching ratios of $H_2$, $A^0$ and $H^\pm$
\cite{Perez:2008ha}. This is because $h_{ij}$ (which determines the strength of the decays mediated by the triplet component) and $v_\Delta$ (which determines the mixing angle) are related by Eq.~(\ref{nu_mass}).

The lower bounds on the masses of the scalars in the HTM from direct searches depend on their decay modes, and such bounds can be quite different from those in the Two Higgs Doublet Model (2HDM).  For example, the decay modes $H^\pm\to W^\pm Z$ and $H_2,A^0\to \nu\nu$ (decays which are not present at tree level in the 2HDM) can be the dominant channels in the HTM \cite{Gunion:1989ci, Datta:1999nc}.  The production processes
$e^+e^-\to \gamma^*, Z^* \to H^{++}H^{--}, H^+H^-$ and $e^+e^-\to Z^*\to A^0H_2$ have rates which depend on gauge couplings. 
The most conservative mass limits which can be imposed are $2M_{H^{\pm\pm}} > M_Z$,  $2M_{H^\pm} > M_Z$ and
$M_{A^0}+M_{H_2} > M_Z$, which ensure that the scalars do not contribute to the width of the $Z$ boson (which is measured very precisely).  Mass bounds for specific decay channels are stronger than these, and can be applied to the scalars in the relevant regions of the HTM parameter space. We will discuss the mass bounds for $H_1$,$H_2$ and $A^0$ below.

The parameters of the scalar potential are also constrained by requiring that it is bounded from below, and the electroweak minimum is a global one. The following constraint on $\lambda_i$ can be derived by requiring that the scalar potential is bounded from below:\footnote{This constraint was also derived in Ref.~\cite{Dey:2008jm} for an alternative parametrization of the scalar potential.}
\begin{equation}
\lambda_1+\lambda_4+2\sqrt{\lambda (\lambda_2 + \lambda_3)}  > 0
\label{bound_from_below}
\end{equation}

An upper limit on $v_\Delta$ can be obtained from considering its effect on the $\rho$ parameter.  In the SM $\rho=1$ at tree level, but higher-order contributions (mainly from virtual top- and bottom-quark loops) give rise to a correction $\delta\rho$, and thus $\rho+\delta\rho\ne 1$.  In the HTM $\delta\rho$ is negative at the {\it tree level}:
\begin{equation}
\rho\equiv 1+\delta\rho={1-{2\epsilon^2\over 1+4\epsilon^2}}\,.
\label{deltarho}
\end{equation}
The measurement $\rho\approx 1$ leads to the bound $v_\Delta/v_0\lsim 0.01$, or $v_\Delta<3$~GeV at $95.5\%$ CL ($2\sigma$) \cite{Abada:2007ux, Fukuyama:2009xk}.  Experimentally, positive values of $\delta\rho$ are preferred ($\rho=1.0004^{+0.0008}_{-0.0004}$ \cite{Amsler:2008zzb}).  However, the above bound on $v_\Delta$ is not rigorous because it is obtained by comparing the above tree-level expression for $\delta\rho$ in the HTM with the experimentally-allowed value of $\delta\rho$, in which the dominant SM contribution from virtual top- and bottom-quark loops has already been computed.  Clearly this is not a consistent treatment of the HTM and SM contributions to $\delta\rho$, which are being evaluated at the tree level and the loop level, respectively.  A full analysis at the loop level in the HTM requires renormalization of $v_\Delta$. Explicit analyses have been performed for a model with a $Y=0$ real scalar triplet, which has significantly fewer scalar fields and does not contain doubly charged Higgs bosons.  The bounds on the triplet vacuum expectation value for the $Y=0$ real scalar triplet are found to be similar in magnitude to those derived from the tree-level analysis \cite{Blank:1997qa}.  We are not aware of an explicit analysis in the HTM, although some studies have been done for other models which contain a $Y=2$ complex scalar triplet, as well as additional fields which are not present in the HTM (e.g., Little Higgs models \cite{Chen:2003fm} and Left-Right symmetric models with $v_\Delta=0$ for the triplet of $SU(2)_L$ \cite{Czakon:1999ga}).  Therefore, in the HTM it seems reasonable to assume a maximum value of the order of a few GeV for the triplet VEV, although the exact bound is not known and will have a dependence on the parameters of the scalar potential.

\subsection{Mass limits for $A^0,H_1,H_2$}

The best limits on the masses of $A^0,H_1,H_2$ come from the CERN LEP experiments.  For scalar masses probed by LEP, the dominant decay modes for $H_2$ and $A^0$ depend on the value of $v_\Delta$ \cite{Perez:2008ha}.  For small triplet VEV ($v_\Delta< 10^{-3}$ GeV) the dominant decay is to two neutrinos ($H_2,A^0\to \nu\nu$), while for larger triplet VEV ($v_\Delta> 10^{-3}$ GeV) the dominant decay is to two $b$ quarks ($H_2,A^0\to b\overline b$) through the doublet component.  The main production mechanisms at LEP are $e^+e^-\to H_1Z$, $e^+e^-\to H_2Z$, $e^+e^-\to H_1A^0$ and $e^+e^-\to H_2A^0$. The relevant couplings are given in Table~\ref{rules}, and they depend on two terms which involve the mixing angles in the CP-even ($\theta_0$) and CP-odd ($\alpha$) sectors.  As explained earlier, in the HTM one always has $\cos\alpha \sim 1$, and $\cos\theta_0\sim 1$ if ${\cal M}^2_{22}\gg {\cal M}^2_{11}$ for the CP-even mass matrix.  In this scenario of $\cos\alpha \sim 1$ and $\cos\theta_0\sim 1$ (which corresponds to most of the parameter space) there will be a SM-like CP-even scalar which can be produced via $e^+e^-\to H_1Z$, and thus the LEP bound $M_{H_1}> 115$ GeV can be applied.  The mechanisms $e^+e^-\to H_2Z$ and $e^+e^-\to H_1A^0$ would have very small cross sections since $\sin\alpha \sim 0$ and $\sin\theta_0\sim 0$.  However, pair production of $A^0$ and $H_2$ is possible via $e^+e^-\to H_2A^0$, since the coupling $ZA^0H_2$ is unsuppressed in this limit.  If the decays modes $H_2\to b\overline b$ and $A^0\to b\overline b$ are dominant then LEP searches can be applied, and the limit $M_{H_2}+M_A > 180$ GeV can be derived \cite{Amsler:2008zzb}.  If the decays $H_2\to \nu\nu$ and $A^0\to \nu\nu$ are dominant, then one can have the signature $e^+e^-\to H_2A^0\to \gamma\nu\nu\nu\nu$, where the photon ($\gamma$) originates from bremsstrahlung from $e^+$ or $e^-$.  Some mass limits can be derived from LEP data for the search for ``$\gamma +$ missing energy'' (the mass bound $M_{H_2}+M_A> 110$ GeV was derived in Ref.~\cite{Datta:1999nc}).  In the case of a large mixing angle ($\cos\theta_0\sim \sin\theta_0\sim 1/\sqrt 2$), all production mechanisms would be relevant. The phenomenology of this scenario is studied in the next section. The couplings $H_iZZ$ and $H_iWW$ are more relevant for phenomenology at the LHC. Pair production of scalars via the $H_1AZ$ and $H_2AZ$ couplings is not so promising at hadron colliders.

\begin{center}
\begin{table}[tb]
\begin{tabular}[t]{|c|c|c|}
\hline
 Vertex & Gauge Coupling  & Approximation
\\
\hline
     $H_2W^+_\mu W^-_\nu$ & $-i{1\over 2} g^2_2 (\sin\theta_0 v_0
-2\cos\theta_0 v_{\Delta}) g_{\mu\nu}$ & $-i{1\over 2} g^2_2({\sqrt{2}\mu
v_0^2\over M_\Delta^2}-2v_\Delta) g_{\mu\nu}$
\\
   $H_2Z_\mu Z_\nu$ & $-i{1\over 2}{g^2_2\over \cos^2\theta_W}(\sin\theta_0
v_0 -4\cos\theta_0 v_{\Delta})g_{\mu\nu} $
& $-i{1\over 2}{g^2_2\over \cos^2\theta_W}
({\sqrt{2}\mu v_0^2\over M_\Delta^2}-4v_{\Delta})g_{\mu\nu}$\\
$AH_2Z_\mu$  & $-{g_2\over
2\cos\theta_W}(\sin\theta_0\sin\alpha+2\cos\alpha\cos\theta_0)(p_1-p_2)_\mu$
&  $-{g_2\over
\cos\theta_W}(p_1-p_2)_\mu$
\\
\hline
$H_1W^+_\mu W^-_\nu$ & $i{1\over 2} g^2_2 (\cos\theta_0 v_0
+2\sin\theta_0 v_{\Delta}) g_{\mu\nu}$ & $i{1\over 2} g^2_2 v_0g_{\mu\nu}$
\\
   $H_1Z_\mu Z_\nu$ & $i{1\over 2}{g^2_2\over \cos^2\theta_W}(\cos\theta_0
v_0 +4\sin\theta_0 v_{\Delta})g_{\mu\nu} $
& $i{1\over 2}{g^2_2\over \cos^2\theta_W}
v_0 g_{\mu\nu} $\\
$AH_1Z_\mu$ & ${g_2\over
2\cos\theta_W}(\cos\theta_0\sin\alpha-2\cos\alpha\sin\theta_0)(p_1-p_2)_\mu$
&  $-{g_2\over \sqrt{2}\cos\theta_W} {\mu v_0 \over M_\Delta^2}(p_1-p_2)_\mu$
\\
\hline
\end{tabular}
\caption{Feynman rules for the CP-even Higgs boson gauge interactions (taken and adapted from \cite{Perez:2008ha}). The momenta are all assumed to be incoming, and $p_1\ (p_2)$ refers to the momentum of the first (second) scalar field
listed in the vertices.  The approximation is based on $v_0\gg v_\Delta,\ M_\Delta \gg M_{H_1}$.}
\label{rules}
\end{table}
\end{center}

\subsection{Discovery channels for $H_1$ and $H_2$ at the LHC}

The phenomenology of the SM Higgs boson [i.e., a scalar which arises solely from an isospin doublet, $\phi^0$ in Eq.~(\ref{SMhiggs})] at the LHC has been studied in great detail, and is reviewed in Ref.~\cite{Djouadi:2005gi}. Much of these analyses can be applied to $H_1$ and $H_2$ of the HTM, whose isospin doublet component $h^0$ corresponds to the SM Higgs scalar multiplied by the mixing angle $\cos\theta_0$ or $\sin\theta_0$ [see Eq.~(\ref{CP-even})].  For an isospin doublet scalar field $h^0$ in the mass range $130 \,{\rm GeV} \to 150$ GeV, the optimal discovery channels are \cite{Djouadi:2005gi}:
\begin{itemize}

\item[{(i)}] Gluon-gluon fusion, followed by decay of $h^0$ to $ZZ^*$: $gg \to h^0$, $h^0\to ZZ^*\to \ell\ell\ell\ell$ 

\item[{(ii)}] Weak-boson fusion, followed by decay of $h^0$ to $\tau^+\tau^-$: $qq\to h^0qq$, $h^0\to \tau^+\tau^-$
 
\item[{(iii)}] Weak-boson fusion, followed by decay of $h^0$ to $W^+W^-$:  $qq\to h^0qq$, $h^0\to W^+W^-$

\end{itemize}
 
The statistical significance of the signal depends on the channel and on the mass of $h^0$, and in channel (iii) it can be as high as $9\sigma$ for ${\cal L}= 30$ fb$^{-1}$.  The significances for channels (i) and (iii) increase as $m_{h^0}$ increases from $130$ GeV to $150$ GeV, while the significance for channel (ii) decreases in the same mass range. We will quote specific numbers for the significances later.  The channel $gg\to h^0\to \gamma\gamma$ is important for the case of $m_{h^0}< 130$ GeV, although the statistical significance is below $5\sigma$ for ${\cal L}= 30$ fb$^{-1}$.

We will not explicitly consider the search channels $gg\to H_1,H_2\to \gamma\gamma$, where the decay $H_1,H_2\to \gamma\gamma$ is mediated by loops involving $W^\pm$, charged fermions and charged scalars. The $W$-loop contribution to $H_1,H_2\to \gamma\gamma$ depends on the couplings $WWH_i$, and in the SM it is the dominant contribution.  The couplings $WWH_i$ depend on two terms, one being proportional to $v_0$ and the other being proportional to $v_\Delta$ (see Table~\ref{rules}). Since $v_0\gg v_\Delta$, the term proportional to $v_0$ is dominant for the case of a large mixing angle $\theta_0\simeq 45^\circ$ of interest to us.  In the case of $M_{H_2}\gg M_{H_1}$, one has $\tan2\theta_0\simeq 4v_ \Delta/v_0$ and the coupling $H_2WW$ is vanishing \cite{Perez:2008ha} because of a cancellation between the two terms -- see the approximate form of the $H_2WW$ coupling in Table~\ref{rules}, where the term in brackets is $\simeq 0$ (from Eq.~\ref{tripletvev}).  The magnitude of the contributions from the loops involving charged fermions is considerably smaller than that of the loop involving $W$.

Charged scalars ($H^{\pm\pm}$ and $H^\pm$) also contribute to the decays $H_1,H_2\to \gamma\gamma$, with the contribution from $H^{\pm\pm}$ having a factor of four enhancement at the amplitude level (because of its electric charge) relative to that from $H^\pm$. The magnitude of these scalar loops depends on the trilinear couplings $H_iH^{++}H^{--}$ and $H_iH^{+}H^{-}$, in which the dominant contribution comes from terms of the form $\lambda_1v_0$ and $\lambda_4v_0$, and there is also a dependence on the mixing angle $\theta_0$.  However, the loop function ($F_0$) for such scalar contributions is much smaller than that for the $W$ boson ($F_1$) (see eg., Ref.~\cite{Djouadi:2005gi}).  For the parameter choice in our numerical analysis ($M_{H_1}\simeq 130$ GeV, $130 \,{\rm GeV} < M_{H_2} < 150 \,{\rm GeV}$ and $M_{H^\pm},M_{H^{\pm\pm}}\simeq 200$ GeV), one has $|F_0|\sim 0.35$ and $|F_1|\sim 8$. Hence the $W$ loop dominates unless large couplings $\lambda_i>1$ are considered.  Importantly, in our numerical analysis we will focus on the phenomenologically interesting case of $0 > \lambda_4 > -1$ and $\lambda_1=0$, and the couplings $H_iH^{++}H^{--}$ have no contribution from $\lambda_4$.  Therefore, in such a scenario the dominant
scalar-loop contribution is from that mediated by $H^\pm$, which does not have the aforementioned enhancement factor of 4. Thus, the branching ratios for $H_{1,2}\to \gamma\gamma$ for the case of $\theta_0\simeq 45^\circ$ in our numerical analysis are essentially the same as that in the SM.

In this work we will focus on the prospects for detection of $H_1$ and
$H_2$ in the above channels (i),(ii) and (iii), for the case of a large
mixing angle $\theta_0$.  For the case of $\cos\theta_0\sim 1$ ($\sin\theta_0\sim 0)$ in the HTM, the eigenstate $H_1$ would be dominantly composed of $h^0$, and thus the above significances can be applied directly to $H_1$. In this scenario, $H_2$ would be almost entirely composed of the triplet field $\Delta^0$, and thus it cannot be produced with an observable rate by the above mechanisms. This can be seen from Table~\ref{rules}, where the $ZZH_2$ and $WWH_2$ couplings (which are needed for weak-boson fusion) are very small. Moreover, when $H_2$ is essentially composed of the triplet field $\Delta^0$ it only couples very weakly to quarks (through its isospin doublet component), thus rendering the gluon-gluon fusion process ineffective. Although the $ZA^0H_2$ coupling is unsuppressed in the limit of $\sin\theta_0\to 0$, the production of scalars via this coupling is not so promising at hadron colliders.   That is, $pp\to Z^* \to A^0H_2$ followed by decays of $H_2$ and $A^0$ to quarks and/or neutrinos does not have such a large cross section, and its experimental signature would suffer from large backgrounds.

\section{Numerical Results}

\begin{figure}[t]
\begin{center}
\includegraphics[origin=c, angle=0, scale=0.51]{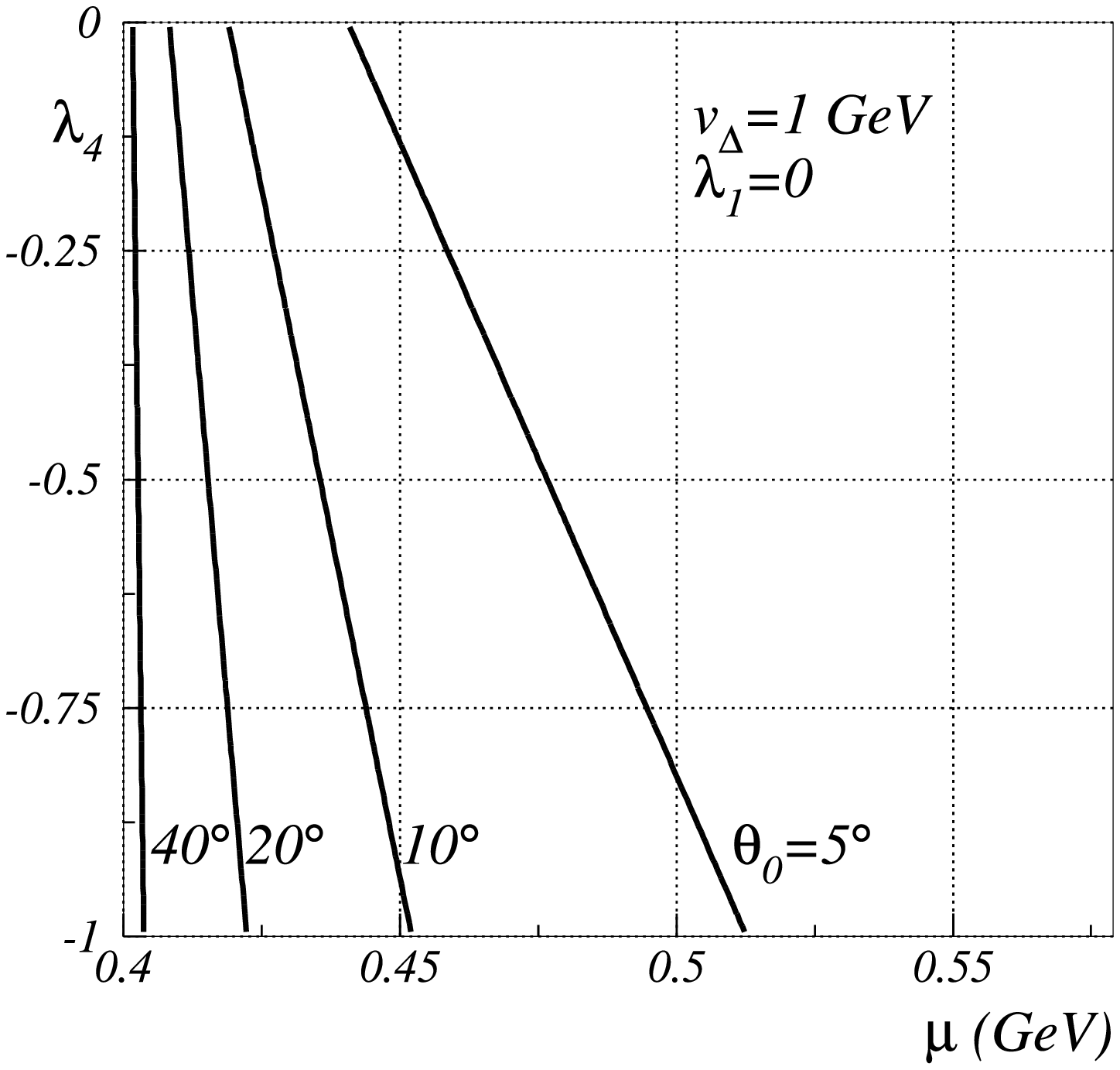}
\includegraphics[origin=c, angle=0, scale=0.51]{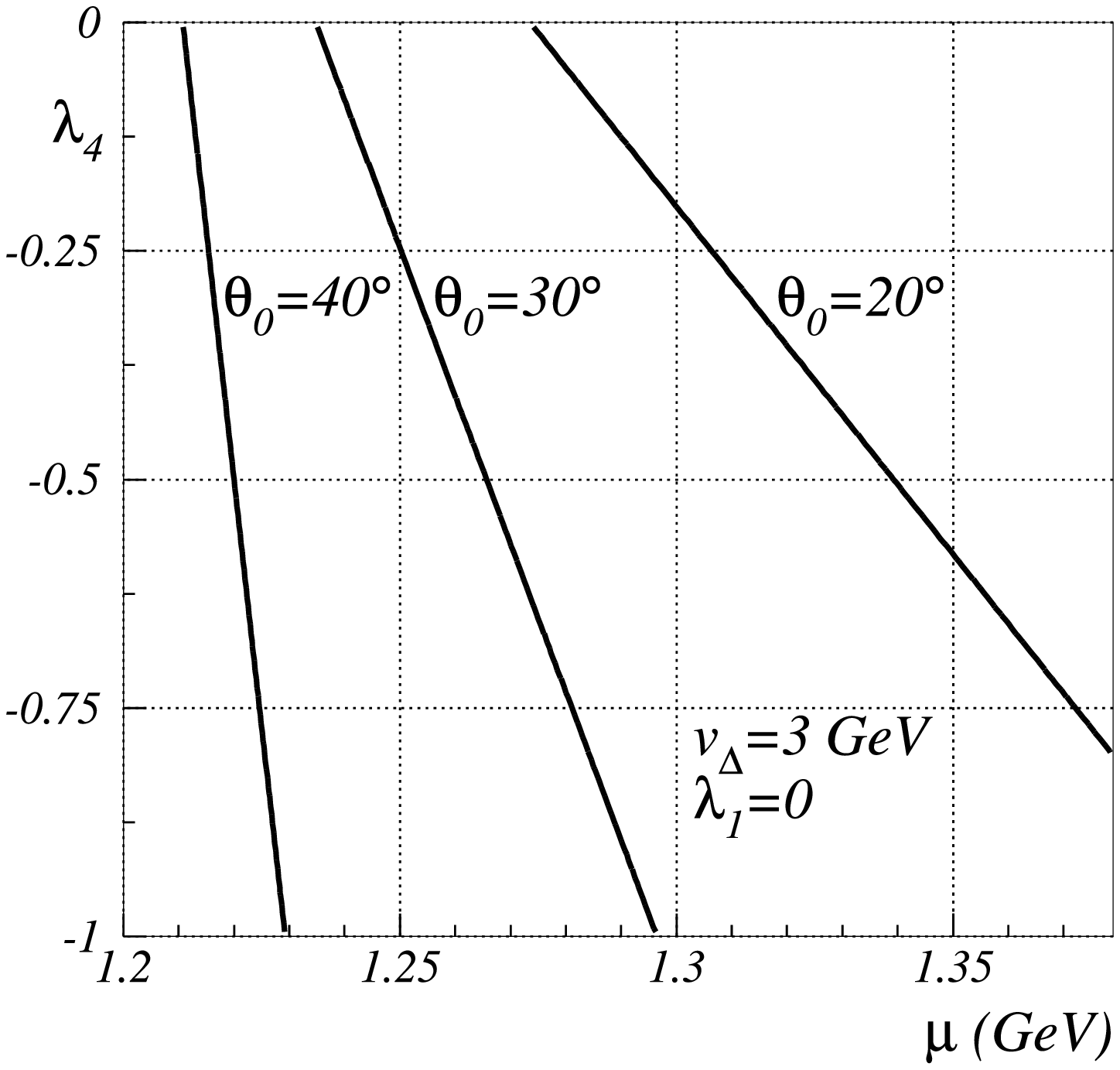}
\vspace*{10mm}
\caption{Contours of the mixing angle $\theta_0$ (in degrees) for the CP-even Higgs bosons, in the $\mu$-$\lambda_4$ plane for $\lambda_4<0$. The left panel has $v_\Delta=1$ GeV and the right panel has $v_\Delta=3$ GeV. In both figures $H_1$ is the lightest Higgs boson in the spectrum and $M_{H_1}\sim 130$ GeV.
Other parameters are fixed as follows: $\lambda\sim 0.566$, $\lambda_1=0$, and $\lambda_{2,3}=1$.}
\label{fig:contour_mix_angle}
\end{center}
\end{figure}

\begin{figure}[t]
\begin{center}
\includegraphics[origin=c, angle=0, scale=0.51]{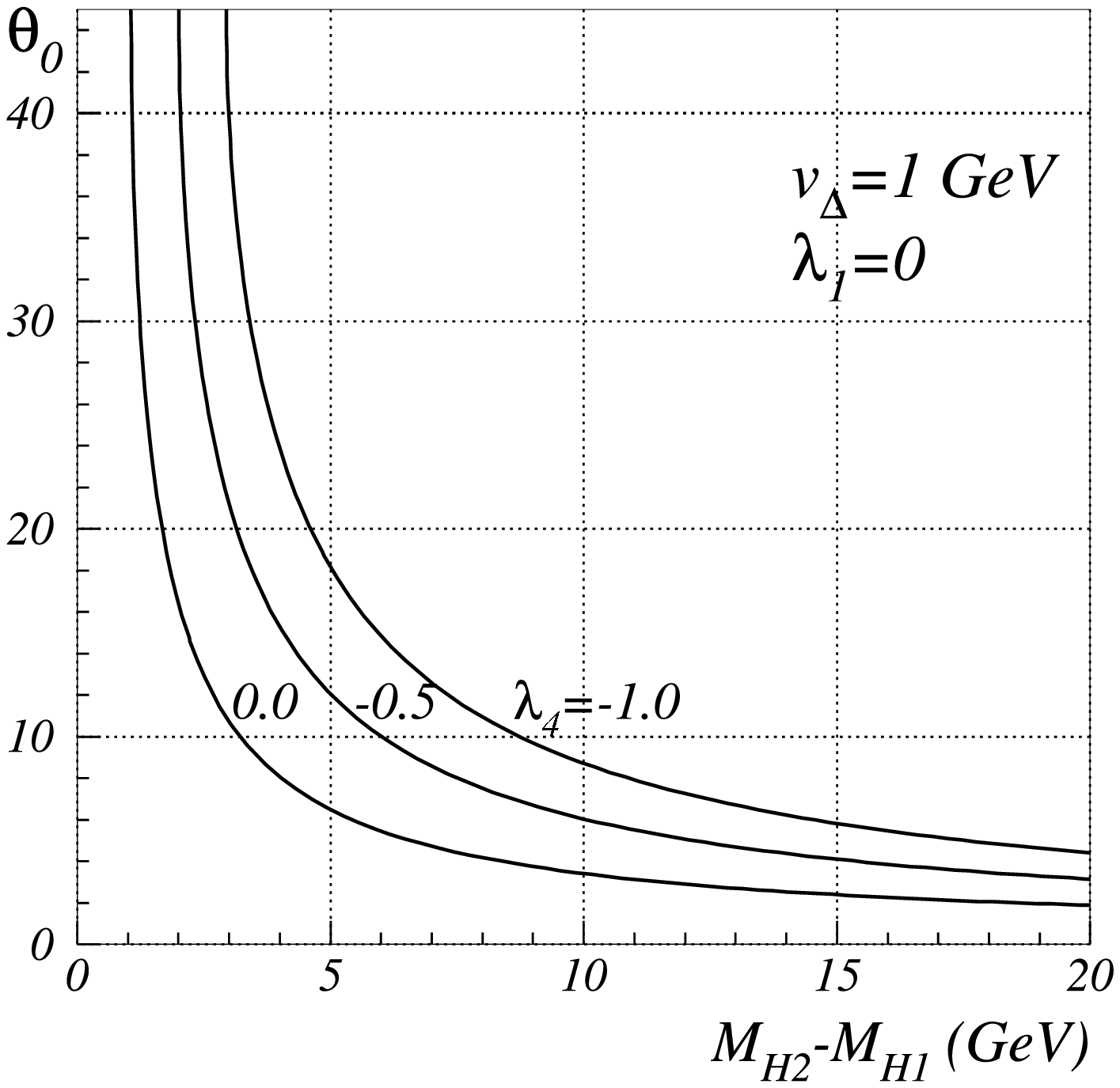}
\includegraphics[origin=c, angle=0, scale=0.51]{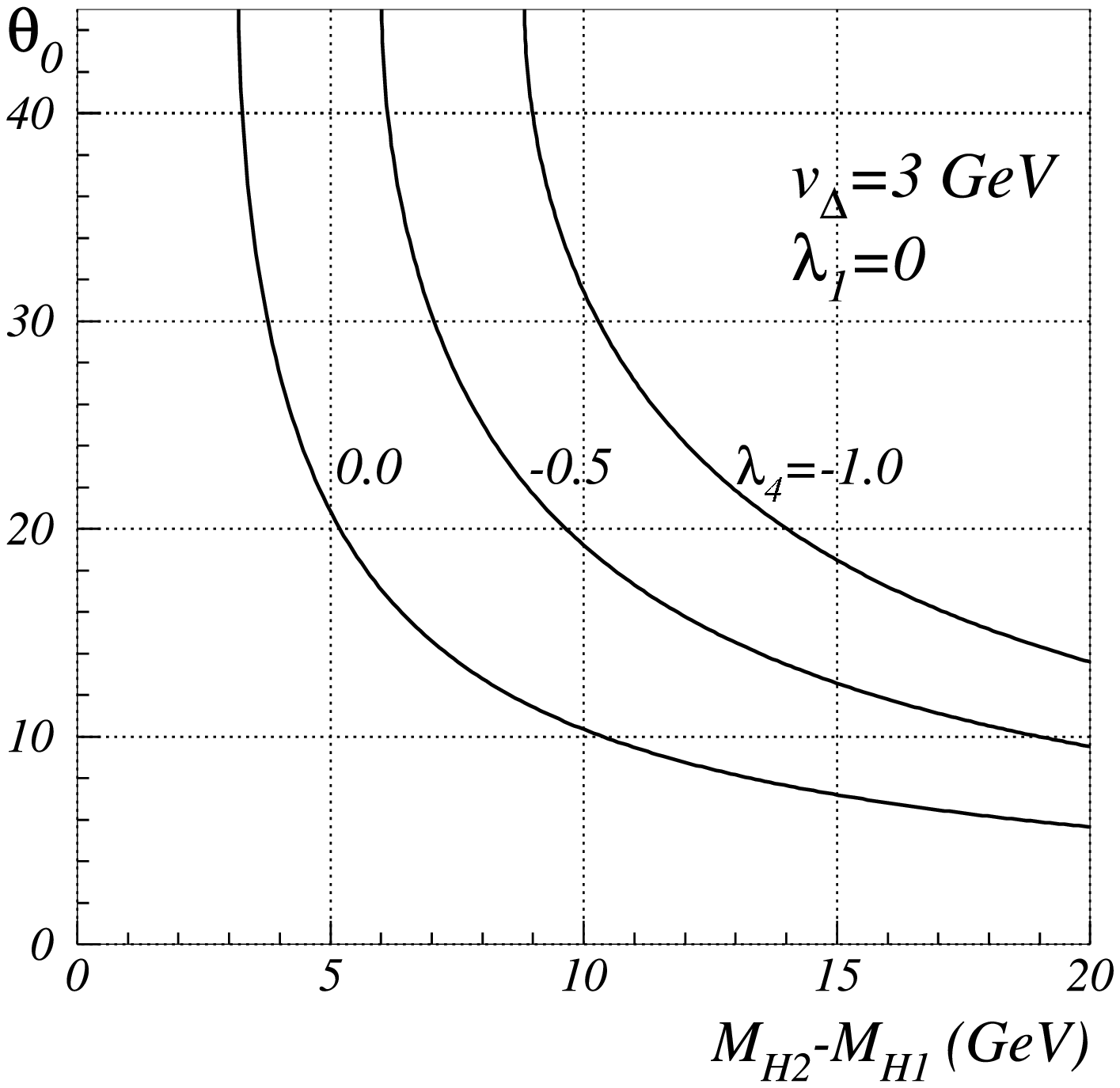}
\vspace*{10mm}
\caption{Mixing angle $\theta_0$ (in degrees) for the CP-even Higgs bosons as a 
function of the mass splitting $M_{H_2}-M_{H_1}$, for $\lambda_4=0,-0.5$ and $-1$.
The left panel has $v_\Delta=1$ GeV and the right panel has $v_\Delta=3$ GeV. 
In both figures $H_1$ is the lightest Higgs boson in the spectrum and $M_{H_1}\sim 130$ GeV.  Other parameters are fixed as follows: $\lambda\sim 0.566$, $\lambda_1=0$, and $\lambda_{2,3}=1$.}
\label{fig:mix_angle_msplit}
\end{center}
\end{figure}

In Ref.~\cite{Dey:2008jm}, the dependence of the mixing angle $\theta_0$ on the theoretical parameter $\mu/v_\Delta$ is studied. In this section we study in detail the parameter space of the HTM where the mixing angle $\theta_0$ can be sizeable.  In Fig.~\ref{fig:contour_mix_angle}, contours of the mixing angle $\theta_0$ for the CP-even Higgs bosons are plotted in the $\mu$-$\lambda_4$ plane for $\lambda_4<0$.  The left panel has $v_\Delta=1$ GeV and the right panel has $v_\Delta=3$ GeV.  Other parameters are fixed as follows: $\lambda\simeq 0.566$, $\lambda_1=0$, and $\lambda_{2,3}=1$.  This choice of parameters satisfies the constraint in Eq.~(\ref{bound_from_below}).  The ratio $\mu/v_\Delta$ is the same (=0.4) in both panels, and it is this ratio which essentially determines the value of $M_{H_2}$ [see Eq.~(\ref{CPeven-masses})].  The choice of $\lambda \simeq 0.566$ gives $M_{H_1}\sim 130$ GeV in both figures, and $H_1$ is the lightest Higgs boson in the spectrum.  The direct search limits for the scalars are respected (see Fig.~\ref{fig:masses}).  The choice of $\lambda\simeq 0.566$ satisfies the condition for maximal mixing\footnote{Note that for fixed $\mu/v_\Delta$ the value of $\lambda$ which gives maximal mixing is slightly different for $v_\Delta=1$ GeV and $v_\Delta=3$ GeV due to the dependence of the second term on $v_\Delta$ in Eq.~(\ref{maxmix}).} in Eq.~(\ref{maxmix}) for $\mu=0.4$ GeV (left panel) and $\mu=1.2$ GeV (right panel), and thus the vertical axis corresponds to the contour of $\theta_0=45^\circ$ (where ${\cal M}^2_{11}={\cal M}^2_{22}$).  For $\mu>0.4$ GeV in the left panel ($\mu>1.2$ in the right panel), one has ${\cal M}^2_{22}>{\cal M}^2_{11}$, and so the mixing angle decreases away from its maximum value (see Eq.~\ref{mixing_angle}).  For a given value of $\mu$, a larger $\theta_0$ can be obtained with a more negative $\lambda_4$.  Negative values $\lambda_4$ enhance the magnitude of the off-diagonal term ${\cal M}^2_{12}$ in the CP-even mass matrix because $\mu$ is taken to be positive (and $\lambda_1=0$).

In Fig.~\ref{fig:mix_angle_msplit}, the mixing angle $\theta_0$ is plotted as a function of the mass splitting $M_{H_2}-M_{H_1}$, for three values of $\lambda_4$.  All other parameters are fixed as in Fig.~\ref{fig:contour_mix_angle}. The only parameter which is varied is $\mu$, starting from $\mu=0.4$ GeV ($\mu=1.2$ GeV) for the left (right) panel, and this generates the mass splitting $M_{H_2}-M_{H_1}$ by increasing $M_{H_2}$ while maintaining $M_{H_1}\sim 130$ GeV. We emphasize that the mass splitting $M_{H_2}-M_{H_1}$ is potentially an experimental observable, and determines whether $H_1$ and $H_2$ can be observed as separate particles.  The left panel has $v_\Delta=1$ GeV and the right panel has $v_\Delta=3$ GeV. All the curves start at $\theta_0=45^\circ$ because the choice of $\lambda \simeq 0.566$ satisfies the condition for maximal mixing in Eq.~(\ref{maxmix}) for $\mu=0.4$ GeV (left panel) and $\mu=1.2$ GeV (right panel). It is evident that maximal mixing can be obtained for mass splittings $M_{H_2}-M_{H_1}\sim 3$ GeV and $\sim 9$ GeV for $v_\Delta=1$ GeV and $v_\Delta=3$ GeV respectively.  For the case of maximal mixing (i.e., ${\cal M}^2_{11}={\cal M}^2_{22}$) the mass splitting is solely caused by the term ${\cal M}^2_{12}$ in Eq.~(\ref{eigenvalues}), and the magnitude of ${\cal M}^2_{12}$ depends on $v_\Delta$.  The region ${\cal M}^2_{22}<{\cal M}^2_{11}$, although possible, is not shown (and would be another line, not necessarily collinear).

The choice of $\lambda_4<0$ increases the observability of $H_2$ at the LHC for two reasons.  Firstly, as can be seen from the mass matrix for the CP-even scalars in Eq.~(\ref{CP-even}), the choice of $\lambda_4<0$ increases the magnitude of the off-diagonal term ${\cal M}^2_{12}$ (because $\mu$ is positive), and hence the mixing angle $\theta_0$ is enhanced.  Secondly, taking $\lambda_4<0$ leads to the mass hierarchy $ M_{{H_2,A^0}} < M_{H^{\pm}}< M_{H^{\pm\pm}}$, and so $H_2$ could be considerably lighter than the charged scalars, as shown in Fig.~\ref{fig:masses}. This latter possibility is usually not emphasized and it is more common to consider the degenerate scenario $M_{H^{\pm\pm}}\sim M_{{H^\pm}} \sim M_{{H_2,A^0}}$, in which the phenomenology of the charged Higgs bosons is likely to be much more important. The fact that $H_2$ could be considerably lighter than the charged scalars means that the detection prospects for $H_2$ could be competitive with those for the charged Higgs bosons, provided that the mixing angle $\theta_0$ is sizeable.
Recently both CDF and D0 Collaboration at the Fermilab Tevatron have searched for the SM Higgs boson and excluded the mass range between 162 GeV and 166 GeV at $95\%$ CL \cite{Aaltonen:2010yv}.
We note that this conclusion may be weakened in the scenario of large mixing in the HTM considered here, provided that the mass splitting between $H_1$ and $H_2$ is of the order of 10 GeV.

As described in Section II-B, the detection prospects for a SM-like Higgs boson at the LHC have been studied in great detail.  We now apply these results to the case of $H_1$ and $H_2$ with a sizeable mixing angle $\theta_0$. For a SM-like Higgs boson $(h^0)$ of mass 130 GeV and with 30 fb$^{-1}$ of integrated luminosity, statistical significances of approximately $5\sigma$, $6\sigma$ and $7\sigma$ can be obtained in the production channels (i) $gg\to h^0, h^0\to ZZ^*$, (ii) $V^*V^*\to h^0, h^0\to \tau\tau$, and (iii) $V^*V^*\to h^0, h^0\to WW^*$, respectively \cite{Djouadi:2005gi}.  For a SM-like Higgs boson of mass 140 GeV, these numbers change to approximately $8\sigma$, $5\sigma$ and $9\sigma$, respectively. To apply these statistical significances to $H_1$ and $H_2$ of a specific mass, one must multiply by the production cross section times branching ratio for $H_{1,2}$ normalized to those for the SM Higgs boson of the same mass.

\begin{figure}[t]
\begin{center}
\includegraphics[origin=c, angle=0, scale=0.51]{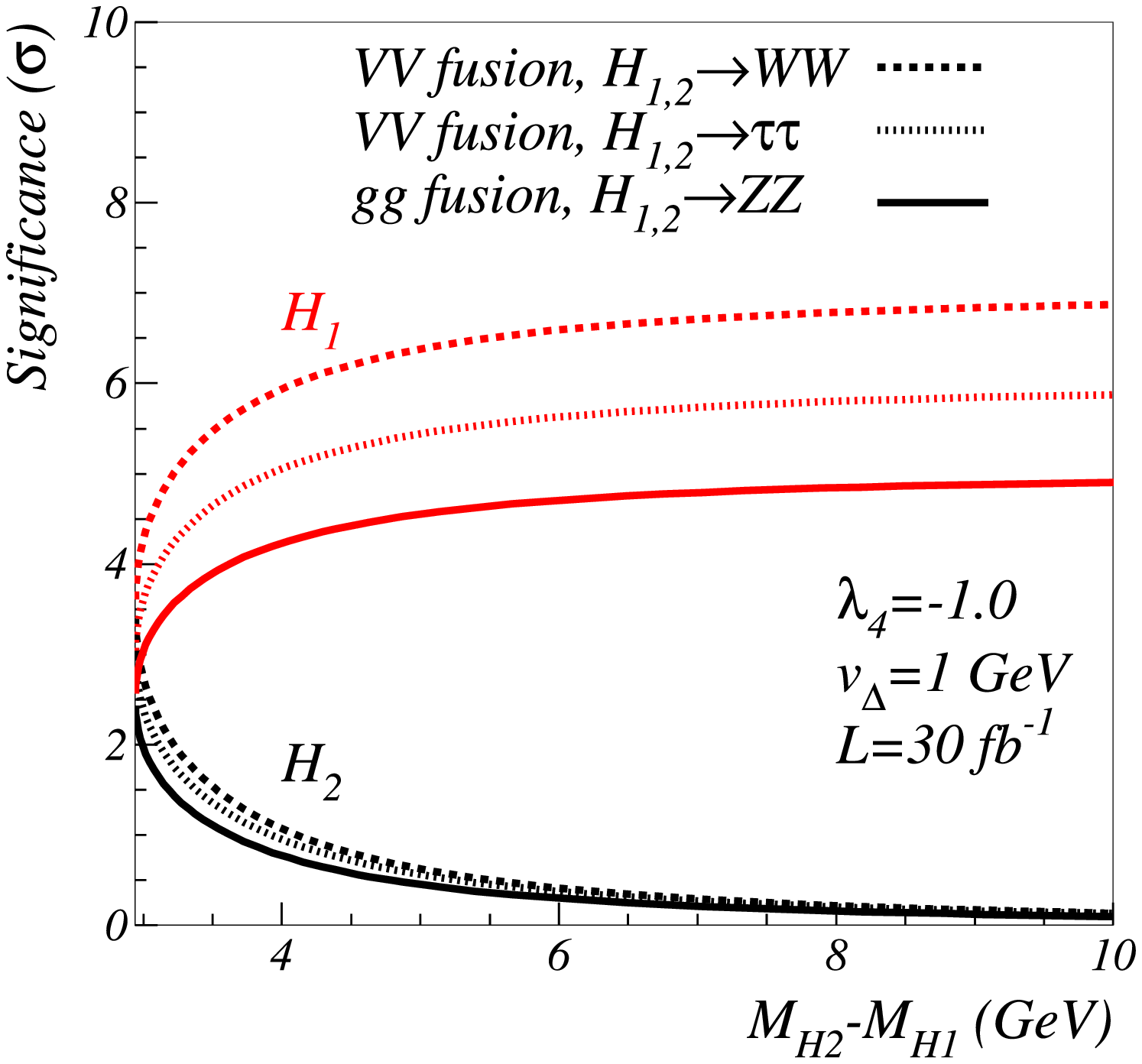}
\includegraphics[origin=c, angle=0, scale=0.51]{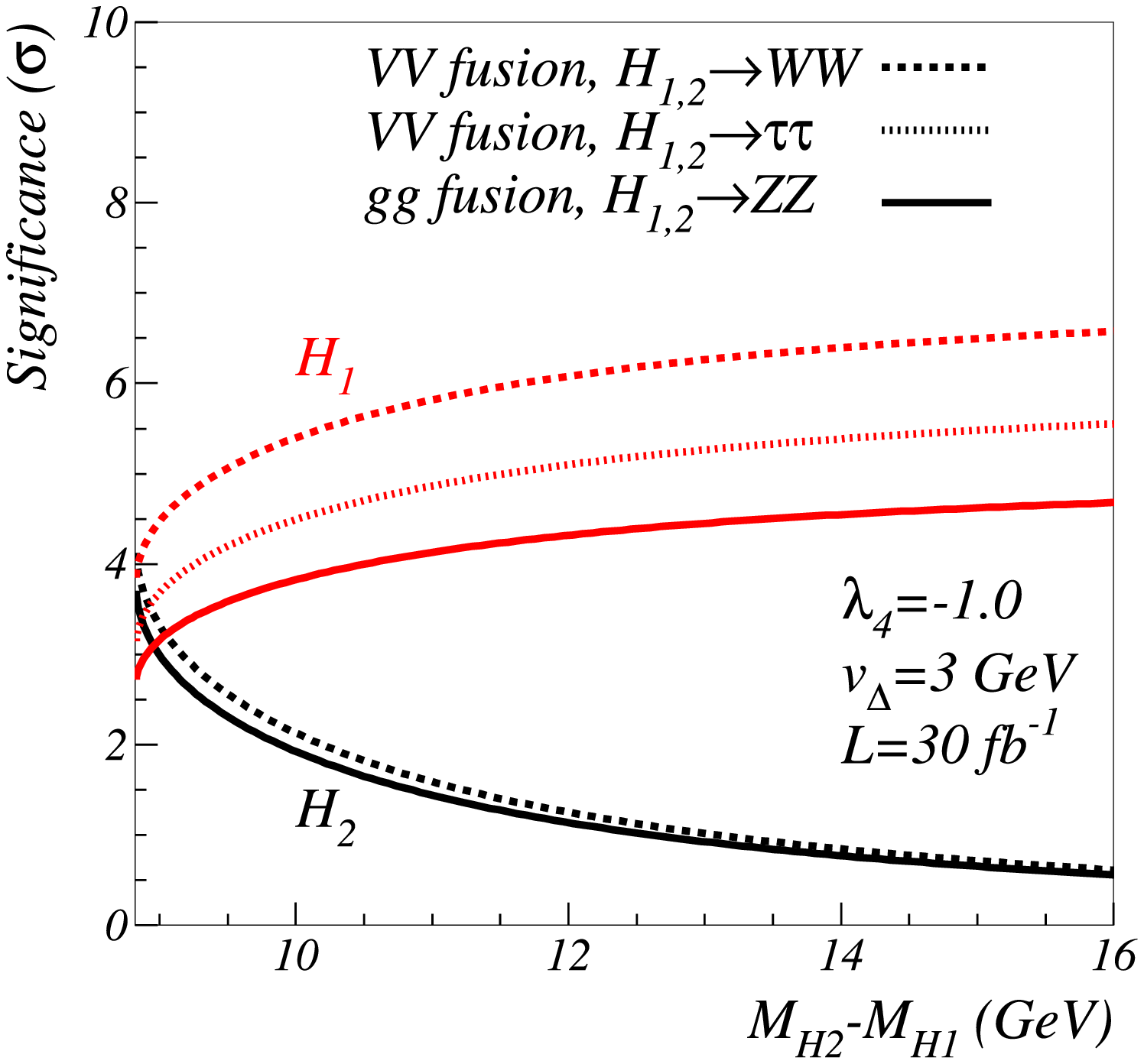}
\vspace*{10mm}
\caption{Statistical significance  for three production mechanisms of $H_1$ and $H_2$ at the LHC with ${\cal L}=30$ fb$^{-1}$.  The left panel has $v_\Delta=1$ GeV and the right panel has  $v_\Delta=3$ GeV. Other parameters are fixed as in Fig.~\ref{fig:mix_angle_msplit}, and $\lambda_4=-1$. In both panels $M_{H_1}\sim 130$ GeV.}
\label{fig:cross_msplit}
\end{center}
\end{figure}

In Fig.~\ref{fig:cross_msplit} the significances for $H_1$ and $H_2$ are plotted as a function of the mass splitting $M_{H_2}-M_{H_1}$ in the three aforementioned channels.  All parameters are fixed to the values used in Fig.~\ref{fig:mix_angle_msplit}, with $\lambda_4=-1$. The left panel has $v_\Delta=1$ GeV and the right panel has $v_\Delta=3$ GeV.  In both panels $M_{H_1}\sim 130$ GeV and so significances of approximately $5\sigma$, $6\sigma$ and $7\sigma$ are used for channels (i), (ii) and (iii) for $H_1$. As $M_{H_2}-M_{H_1}$ increases, one can see that the statistical signficance for $H_1$ in all three channels approaches that of a SM-like Higgs boson. For smaller $M_{H_2}-M_{H_1}$ (corresponding to large $\sin\theta_0$) the significance for $H_1$ is below that of a SM-like Higgs of the same mass.  For the three channels for $H_2$, the mass range is $133\, {\rm GeV} < M_{H_2} < 140$ GeV in the left panel, and $138 \,{\rm GeV} < M_{H_2} < 146$ GeV in the right panel.  In the left panel we use the SM significances of $5\sigma$, $6\sigma$ and $7\sigma$ (corresponding to a SM-like Higgs boson of mass 130 GeV) and show all three channels for $H_2$.  In the right panel we use the SM significances of $8\sigma$ and $9\sigma$ for channels (i) and (iii), respectively, (corresponding to a SM-like Higgs boson of mass 140 GeV) over the displayed mass range $138 \,{\rm GeV} < M_{H_2} < 146$ GeV. Note that these values slightly underestimate the significances for $M_{H_2}>140$ GeV in the right panel, because a SM-like Higgs boson of mass 146 GeV would have a significance closer to $9\sigma$ and $10\sigma$ for channels (i) and (iii), respectively.  We omit channel (ii) for $H_2$ in the right panel, whose significance is considerably lower than that in channels (i) and (iii), and falls from $5\sigma$ to $3\sigma$ for a SM-like Higgs boson with a mass between 140 GeV and 146 GeV.

In the most optimistic scenario (corresponding to $\theta_0=45^\circ$) $H_2$ can be produced with a cross section about half that of the SM Higgs boson, and thus significances of up to $4\sigma$ could be obtained for $H_2$ with 30 fb$^{-1}$ at the LHC when $M_{H_1} = 130$ GeV.  With 300 fb$^{-1}$ the significance would increase to $12\sigma$, and in the right panel even $3\sigma$ evidence would be possible for $M_{H_2}-M_{H_1}=15$ GeV.  For larger values of $M_{H_1}$ the significance can be greater.  If $M_{H_1} = 160$ GeV and the mass splitting is $\sim 10$ GeV, a significance of $5\sigma$ can be achieved for both $H_1$ and $H_2$ with an integrated luminosity of 7 fb$^{-1}$.  The largest significance is for $M_{H_{1,2}} > 200$ GeV (as much as $40\sigma$ for a SM-like Higgs boson with 100 fb$^{-1}$), because $H_1$ and $H_2$ can decay to two on-shell $Z$ bosons with a large branching ratio.  Therefore $H_2$ can be produced with a sizeable rate in the search channels for the SM Higgs boson, provided that the mass splitting $M_{H_2}-M_{H_1}$ is sufficiently small, as described above.  For larger mass splittings, one has the well-known result that $H_2$ is difficult to observe at the LHC because the cross section is much smaller ($< 0.1$) than that for a SM Higgs boson of the same mass.

An important issue is whether the individual signals for $H_1$ and $H_2$ can be separated.  If not, then the case of $H_1$ and $H_2$ being almost degenerate in mass would have a signature indistinguishable from that of one (SM-like) Higgs boson. Here the channel $gg \to H_1,H_2\to ZZ^*\to \ell\ell\ell\ell$ is of particular importance because an accurate measurement of the masses $M_{H_1}$ and $M_{H_2}$ can be achieved. For an integrated luminosity of 300 fb$^{-1}$ the precision is of the order of $0.1$ GeV for a SM-like Higgs boson with a mass of less than $400$ GeV \cite{Djouadi:2005gi}. For 30 fb$^{-1}$ one might have a resolution of GeV order, and so it is possible to separate the signals for $gg \to H_1\to ZZ^*\to \ell\ell\ell\ell$ and $gg \to H_2\to ZZ^*\to \ell\ell\ell\ell$, as long as the mass splitting $M_{H_2}-M_{H_1}$ is within a few GeV. Therefore, the channels $gg \to H_1,H_2\to ZZ^*\to \ell\ell\ell\ell$ have the potential to disentangle the signals for $H_1$ and $H_2$, which is crucial in order to confirm that the signal has originated from two distinct scalars, $H_1$ and $H_2$.  In contrast, for the weak-boson fusion channels it is unlikely that the signals for $H_1$ and $H_2$ can be resolved.  This is because the signal would be an excess of events above the background, and the mass of $H_1$ and $H_2$ cannot be measured very well. Therefore in the weak-boson fusion channels the signal for $H_1$ and $H_2$ would be indistinguishable from that of the SM Higgs boson.

We note that maximal mixing is also possible for values of $v_\Delta$ much less than
1 GeV. However, in this case ${\cal M}^2_{12}\ll {\cal M}^2_{11}\sim{\cal M}^2_{22}$, 
and a large mixing angle is only possible for a tiny mass splitting $M_{H_2}-M_{H_1}$. For 
$v_\Delta$ below the MeV scale, the channel $H_2\to \nu\nu$ becomes an important
decay mode. Therefore, in this scenario with a large mixing angle
(which requires $H_1$ and $H_2$ to have essentially the same mass), $H_2$ can be produced
by the above production mechanisms followed by the invisible decay $H_2\to \nu\nu$.

We now comment on the phenomenology of the charged Higgs bosons ($H^{\pm\pm},H^\pm$) in this scenario of a large mixing angle for the CP-even scalar sector.  When the triplet VEV is of the order of 1 GeV (which is the case in our numerical analysis), the dominant decay modes of the charged scalars are $H^{\pm\pm}\to W^\pm W^\pm$ and $H^\pm\to W^\pm Z, tb$ (e.g., see \cite{Perez:2008ha}). We note that detection prospects for these decay modes are not as promising as those for $H^{\pm\pm}\to \ell^\pm\ell^\pm$ and $H^{\pm}\to \ell^\pm\nu$ (which dominate for smaller values of the triplet VEV).  Ref.~\cite{Han:2007bk} studies the observability of the channel $pp\to H^{++}H^{--}\to W^+W^+W^-W^-$ at the LHC and shows that a signal cross section of around $0.12$ fb can be obtained for $M_{H^{\pm\pm}}=300$ GeV, with a background of $0.12$ fb. This result suggests that the significance in this channel with an integrated luminosity of 30 fb$^{-1}$ is similar to that for $H_2$ when the mixing angle $\theta_0$ is sizeable.  The cross section for the production mechanisms $pp\to W^{\pm *}\to W^\mp H^{\pm\pm}$ and fusion via $W^{\pm *} W^{\pm *} \to H^{\pm\pm}$ \cite{Huitu:1996su, Vega:1989tt} depends on $v_\Delta$. Explicit simulations \cite{Azuelos:2004dm} suggest that detection of $H^{\pm\pm}\to W^\pm W^\pm$ in these channels is difficult for $v_\Delta\sim 1$ GeV.  Importantly, as shown in Fig.~\ref{fig:masses}, for $\lambda_4<0$ one has the mass hierarchy $M_{H^{\pm\pm}}> M_{{H^\pm}} > M_{{H_2,A^0}}$, and so $H_2$ is lighter than the charged scalars.  Therefore the detection prospects for $H_2$ might be comparable (or even better) than those for $H^{\pm\pm}$ and $H^\pm$ if the mixing angle in the CP-even sector is sizeable and if $v_\Delta$ is of the order of 1 GeV.

\section{Summary}

The Higgs Triplet Model (HTM) contains two CP-even Higgs bosons ($H_1$ and $H_2$).  Most previous studies assume that the heavier one ($H_2$) is mostly the neutral component of the Higgs triplet and interacts very weakly with the standard model (SM) particles due to a vanishing mixing with the Higgs doublet.  Such a scenario makes it difficult to detect $H_2$ at colliders.  By studying the most general Higgs potential in the HTM that is invariant under the $SU(2) \otimes U(1)_Y$ symmetry, we have found the condition for the mixing between the CP-even scalar fields to be maximal and shown that this occurs when their mass eigenstates are almost degenerate.  More specifically, the parameter $M_{\Delta}$ that controls the overall mass scale of the Higgs triplet fields should be brought down to $\sim$ TeV.  This is phenomenologically very interesting because various Higgs bosons in the model can be accessible at the CERN LHC.   More importantly, in the large mixing scenario, both $H_1$ and $H_2$ can be looked for through the search channels for the SM Higgs boson.

In this paper, we take the Higgs triplet vacuum expectation value (VEV) $v_\Delta \sim {\cal O}(1 {\rm GeV})$, as constrained by the experimental $\rho$ parameter.  In this case, the detection prospects of the charged Higgs bosons ($H^{\pm}$ and $H^{\pm\pm}$) are less promising than 
at small values of $v_\Delta$.  We then select the parameter space by requiring that the mass of lighter CP-even Higgs boson $M_{H_1} \sim 130$ GeV, that the Higgs potential is bounded from below, and that the electroweak minimum is a global one.  We show the dependence of the mixing angle $\theta_0$ of the CP-even Higgs bosons on the model parameters $\mu$ and $\lambda_4$, with the former controlling the mass splitting between $H_1$ and $H_2$ and the latter controlling the mass splitting between $H_2$, $H^{\pm}$ and $H^{\pm\pm}$.  We also study the dependence of $\theta_0$ on the mass splitting $M_{H_2} - M_{H_1}$.  Finally, we examine the statistical significance of the two CP-even Higgs bosons at the LHC with an integrated luminosity of 30 fb$^{-1}$.  For a sufficiently small mass splitting, depending on the value of $v_\Delta$, the significance of $H_2$ can reach up to $4\sigma$ for the maximal mixing scheme.  We also note that the situation becomes better when $M_{H_1}$ is larger.

\vspace{1cm}
{\it Acknowledgments}: We thank J.~S.~Lee and L.~F.~Li for useful discussions.  This work was financially supported in part by the National Science Council of Taiwan, R.~O.~C.\ under Grant No.~NSC~97-2112-M-008-002-MY3 and the NCTS.  A.G.A is supported by the ``National Central University Plan to Develop First-class Universities and Top-level Research Centers.''

\end{document}